\documentclass[]{research}

\usepackage{algpseudocode}
\usepackage[ruled,vlined]{algorithm2e}

\usepackage{amsmath}
\usepackage{amsfonts}
\usepackage{hyperref}
\usepackage{url}
\usepackage{booktabs}
\usepackage{multirow}
\usepackage{pgfplots}
\usepackage{graphicx}
\usepackage{array}
\pgfplotsset{compat=1.18}
\usepackage{arydshln}

\usepackage{xspace}

\newcolumntype{L}[1]{>{\raggedright\let\newline\\\arraybackslash\hspace{0pt}}m{#1}}
\newcolumntype{C}[1]{>{\centering\let\newline\\\arraybackslash\hspace{0pt}}m{#1}}
\newcolumntype{R}[1]{>{\raggedleft\let\newline\\\arraybackslash\hspace{0pt}}m{#1}}

\newcommand{\sects}[1]{\S~\ref{sect:#1}}

\newcommand{\fig}[1]{Figure~\ref{fig:#1}}

\newcommand{\lblsect}[1]{\label{sect:#1}}

\newcommand{\ignorethis}[1]{}

\makeatletter
\DeclareRobustCommand\onedot{\futurelet\@let@token\@onedot}
\def\@onedot{\ifx\@let@token.\else.\null\fi\xspace}

\makeatother

\definecolor{citecolor}{rgb}{34,139,34}
\definecolor{mydarkblue}{rgb}{0,0.08,1}
\definecolor{mydarkgreen}{rgb}{0.12,0.7,0.12}
\definecolor{mydarkred}{rgb}{0.8,0.02,0.02}
\definecolor{mydarkorange}{rgb}{0.40,0.2,0.02}
\definecolor{mypurple}{RGB}{111,0,255}
\definecolor{myred}{rgb}{1.0,0.0,0.0}
\definecolor{mygold}{rgb}{0.75,0.6,0.12}
\definecolor{mydarkgray}{rgb}{0.66,0.66,0.66}

\newcommand{\mypara}[1]{\vspace{0pt}\noindent\textbf{#1}}

\definecolor{darkgreen}{rgb}{0.15, 0.75, 0.15}
\definecolor{mitblue}{rgb}{0.88,0.95,0.96}
\definecolor{lightblue}{rgb}{0.90, 0.95, 0.99}

\newcommand{\kmeans}{{\emph{k}-means}\xspace}
\newcommand{\method}{flash-kmeans\xspace}
\newcommand{\Method}{Flash-kmeans\xspace}

\newcommand{\FA}{FlashAssign\xspace}

\newcommand{\siu}{{sort-inverse update}\xspace}
\newcommand{\Siu}{{Sort-inverse update}\xspace}
\newcommand{\SIU}{{Sort-Inverse Update}\xspace}

\newcommand{\cso}{{chunked stream overlap}\xspace}

\newcommand{\Compileheuristic}{{Cache-aware compile heuristic}\xspace}
\newcommand{\compileheuristic}{{cache-aware compile heuristic}\xspace}
\newcommand{\ttfr}{{time-to-first-run}\xspace}

\makeatletter
\def\adl@drawiv#1#2#3{%
        \hskip.5\tabcolsep
        \xleaders#3{#2.5\@tempdimb #1{1}#2.5\@tempdimb}%
                #2\z@ plus1fil minus1fil\relax
        \hskip.5\tabcolsep}
\newcommand{\cdashlinelr}[1]{%
  \noalign{\vskip\aboverulesep
           \global\let\@dashdrawstore\adl@draw
           \global\let\adl@draw\adl@drawiv}
  \cdashline{#1}
  \noalign{\global\let\adl@draw\@dashdrawstore
           \vskip\belowrulesep}}
\makeatother

\title{Flash-KMeans: Fast and Memory-Efficient Exact K-Means}
\author[*]{Shuo Yang}
\author[*]{Haocheng Xi}
\author[]{Yilong Zhao}
\author[]{Muyang Li}
\author[]{Xiaoze Fan}
\author[]{Jintao Zhang}
\author[]{Han Cai}
\author[]{Yujun Lin}
\author[]{Xiuyu Li}
\author[]{Kurt Keutzer}
\author[]{Song Han}
\author[]{Chenfeng Xu}
\author[]{Ion Stoica}

\affiliation[]{}

\contribution[*]{Equal Contribution}

\abstract{

\kmeans has historically been positioned primarily as an offline processing primitive, typically used for dataset organization or embedding preprocessing rather than as a first-class component in online systems. In this work, we revisit this classical algorithm under the lens of modern AI system design and enable \kmeans as an online primitive.
We point out that existing GPU implementations of \kmeans remain fundamentally bottlenecked by low-level system constraints rather than theoretical algorithmic complexity. Specifically, the assignment stage suffers from a severe IO bottleneck due to the massive explicit materialization of the $N \times K$ distance matrix in High Bandwidth Memory (HBM). Simultaneously, the centroid update stage is heavily penalized by hardware-level atomic write contention caused by irregular, scatter-style token aggregations. 
To bridge this performance gap, we propose \method, an IO-aware and contention-free \kmeans implementation for modern GPU workloads. \Method introduces two core kernel-level innovations: (1) \textit{\FA}, which fuses distance computation with an online argmin to completely bypass intermediate memory materialization; (2) \textit{\siu}, which explicitly constructs an inverse mapping to transform high-contention atomic scatters into high-bandwidth, segment-level localized reductions. Furthermore, we integrate algorithm-system co-designs, including \cso and \compileheuristic, to ensure practical deployability.
Extensive evaluations on NVIDIA H200 GPUs demonstrate that \method achieves up to \textbf{17.9$\times$} end-to-end speedup over best baselines, while outperforming industry-standard libraries like \texttt{cuML} and \texttt{FAISS} by \textbf{33$\times$} and over \textbf{200$\times$}, respectively. 
At the kernel level, \FA and \SIU deliver up to \textbf{21.2$\times$} and \textbf{6.3$\times$} speedups. Beyond raw compute, our pipelined co-design enables seamless out-of-core execution on up to \textbf{one billion points} with a \textbf{10.5$\times$} speedup, and our heuristics slash configuration tuning overhead by \textbf{175$\times$} with near-zero performance degradation. 
By systematically restructuring execution around underlying hardware constraints, \method delivers mathematically exact, scalable, and highly deployable acceleration across diverse AI workloads.}

\correspondence{Shuo Yang at \href{mailto:andy_yang@berkeley.edu}{andy\_yang@berkeley.edu}, Chenfeng Xu at \href{mailto:xuchenfeng@utexas.edu}{xuchenfeng@utexas.edu}}

\metadata[Code]{\url{https://github.com/svg-project/flash-kmeans}}

\begin{document}

\maketitle

\section{Introduction}


\begin{figure}[t]
    \centering
    \includegraphics[width=0.8\textwidth]{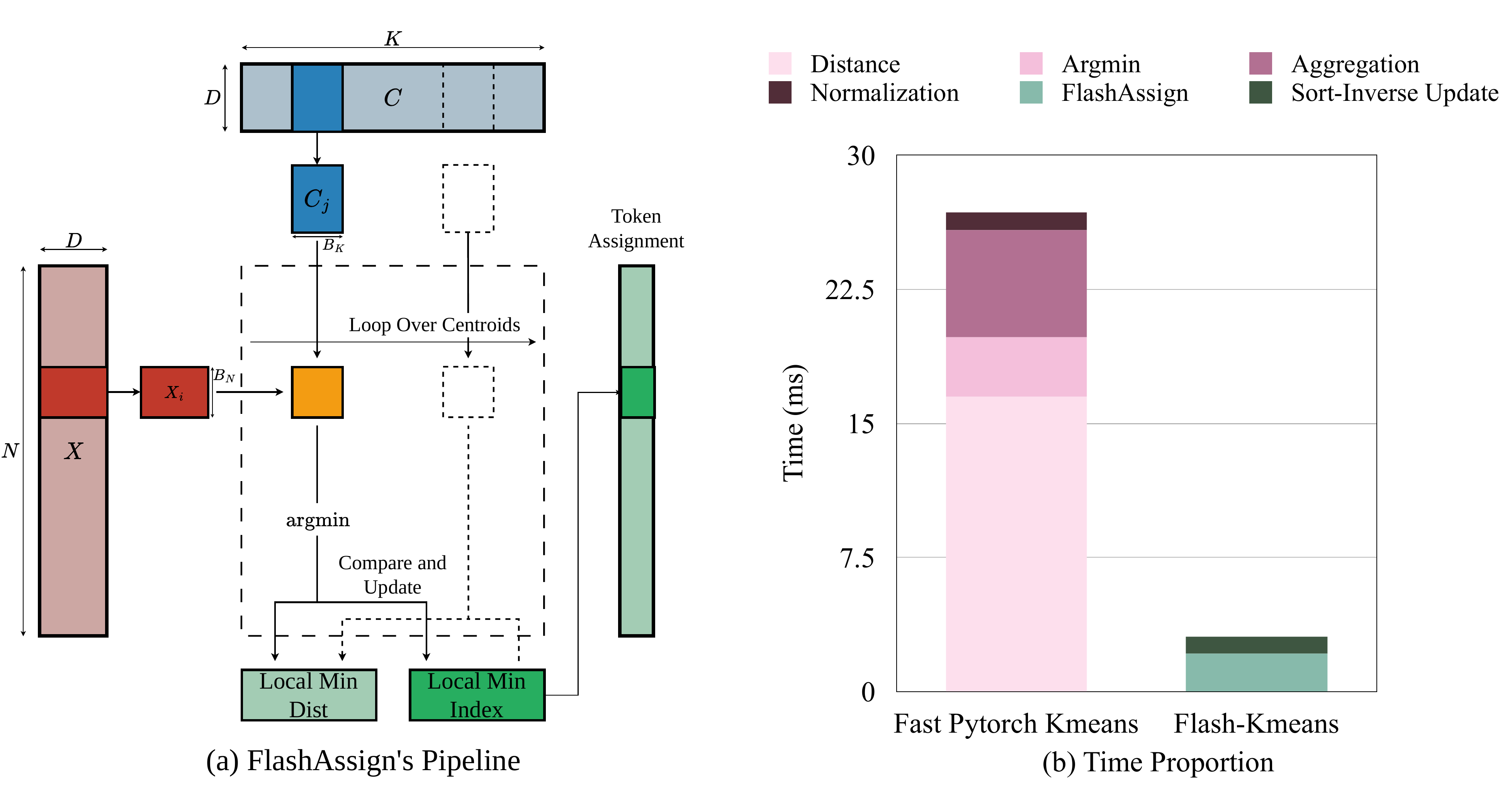} 
    \caption{Overview of \method and Performance Breakdown.  
    (a) Inspired by IO-aware attention mechanisms, \FA streams data blocks from HBM to SRAM, fusing distance computation with an online argmin operator to completely bypass the materialization of the massive $N \times K$ distance matrix. 
    (b) Compared to standard \kmeans implementations, \method drastically compresses both the assignment IO bottleneck and the update synchronization bottleneck.}
    \label{fig:flash_kmeans_overview}
\end{figure}

\kmeans \cite{1056489, mcqueen1967smc} is one of the most classical and widely used clustering algorithms, and has long been treated as a mature, general-purpose component. 
In traditional settings, \kmeans typically appears as part of offline data processing pipelines \cite{10.1145/1772690.1772862}, and most optimizations focus on the algorithmic level, such as faster convergence, fewer distance evaluations, or better approximations \cite{10.5555/3041838.3041857, pmlr-v37-ding15, bachem2018scalablekmeans}. 
Although these algorithmic improvements successfully reduce theoretical FLOPs, many of them fail to translate into real end-to-end speedups on modern GPUs. 
This discrepancy arises because traditional optimizations are often incompatible with modern hardware design principles: dense compute (e.g., matrix multiplications via Tensor Cores) is exceedingly cheap \cite{a100}, while frequent, fragmented memory operations incur massive overheads. Consequently, lowering FLOPs does not necessarily correlate with wall-clock time reduction, and heavily overlooks the dominant constraints of memory bandwidth and data movement \cite{dao2022flashattention,dao2023flashattention2}.

This performance gap caused by implementation-level factors is significantly amplified in modern AI workloads. \kmeans usage is undergoing a clear paradigm shift, evolving from an offline analytics tool into a high-frequency, online operator within training and inference pipelines (e.g., vector quantization and sparse routing) \cite{roy2020routingtransformer, zhu2025tactic, liu2025clusterkvm}. 
This shift manifests in three dimensions: (1) execution moves from CPUs to GPUs, creating clear potential to exploit hardware accelerators; (2) usage moves from offline to online invocation, making low latency per invocation strictly essential; and (3) computation moves to longer sequences and larger batches, making throughput and scalability crucial requirements. 

To understand why standard implementations fail to meet these new requirements, we must first define the \kmeans workload. A standard Lloyd's iteration consists of two distinct stages: an \textit{assignment stage}, which computes the distance between every data point and all centroids to find the nearest cluster, and a \textit{centroid update stage}, which is essentially an averaging operation that aggregates the features of all points assigned to each specific cluster to form the new centroids. 

When profiling this standard pipeline on modern GPUs \cite{raschka2020machine, johnson2019billion, fastkmeans2025}, we find that the end-to-end performance is severely bottlenecked by kernel-level dataflow and write-path serialization. Concretely, this performance gap is driven by three primary challenges:

\begin{enumerate}
    \item \textbf{IO-bound assignment in assignment stage:} Consider a workload consisting of $N$ data points in $d$-dimensional space, clustered into $K$ centroids. In each iteration, the assignment step compares every point against all $K$ centroids, resulting in a computational cost of $O(NKd)$. 
    Standard implementations execute this by first computing distances and then applying $\arg\min$, which explicitly materializes a massive distance matrix $D\in\mathbb{R}^{N\times K}$. Writing $D$ to High Bandwidth Memory (HBM) and immediately reading it back incurs immense memory traffic. Under a representative setting with $N=65536, K=1024, d=128, B=32$, the distance computation itself takes only 2.6 milliseconds, whereas the memory time dominated by materializing and consuming $D$ takes about 23 milliseconds.
    
    \item \textbf{Atomic write contention in update stage:} The centroid update stage requires aggregating data grouped by cluster indices. Standard implementations typically perform scatter-style updates, where each thread atomically accumulates its point's data into a shared sum and count buffer based on its cluster id. Consequently, many threads concurrently attempt to update the same centroid (especially for unbalanced, "hot" clusters), causing severe atomic contention and hardware-level serialization. On an NVIDIA H200 GPU, we measure only 50~GB/s effective bandwidth for this stage, drastically below the achievable bandwidth of regular reductions.
    
    \item \textbf{System-level constraints:} Beyond kernel-level bottlenecks, real-world deployments face severe system constraints. When the input batch cannot reside entirely in VRAM, chunk-wise execution introduces heavy host-to-device communication overheads. Furthermore, modern AI workloads often exhibit dynamic shapes, which amplify compilation and configuration tuning costs, drastically increasing the \ttfr and hurting practical usability.
\end{enumerate}

To bridge this performance gap, we propose \method, an efficient implementation of \kmeans designed for modern GPU workloads. 
\Method does not alter the mathematical formulation of the standard Lloyd \kmeans, nor does it introduce approximations. Instead, it restructures the assignment stage, the centroid update stage, and the system execution path around the three hardware bottlenecks discussed above. 

Concretely, we introduce three key techniques. First, to eliminate distance materialization, we propose \textit{\FA} (\fig{flash_kmeans_overview}a), which fuses streaming distance computation with an online argmin procedure. This avoids constructing the $N\times K$ intermediate matrix entirely, thereby eliminating the huge related overhead \cite{dao2022flashattention}. 
Second, to resolve atomic contention, we introduce \textit{\siu} (\fig{siu_mechanism}b), which explicitly sorts the assignment vector by cluster id and performs aggregations in this sorted logical order. This rewrites high-contention, per-token atomic scatters into highly regular, segment-level localized merges. 
Finally, we introduce a set of \textit{algorithm-system co-design} optimizations, including \cso to hide PCIe communication overhead for large-scale execution, and a \compileheuristic to reliably match near-optimal configurations without expensive tuning, solving the dynamic deployment challenge.

We prototype \method with customized GPU kernels and systematically evaluate it on an NVIDIA H200 GPU. 
We first conduct a kernel-level breakdown, revealing that \FA speeds up the assignment kernel by up to \textbf{21.2$\times$}, and \SIU accelerates the centroid update kernel by up to \textbf{6.3$\times$}. 
Scaling to end-to-end iterations, we sweep across various numbers of points $N$, clusters $K$, feature dimensions $d$, and batch sizes $B$. \method consistently achieves state-of-the-art performance across all representative workload regimes. As shown in \fig{flash_kmeans_overview}b, it outperforms the strongest baseline by up to \textbf{17.9$\times$}, and delivers massive speedups of up to \textbf{33$\times$} and over \textbf{200$\times$} against widely adopted industry-standard libraries like NVIDIA \texttt{cuML} \cite{raschka2020machine} and \texttt{FAISS} \cite{johnson2019billion}, respectively.
Beyond raw kernel speedups, our algorithm-system co-design optimizations prove critical for real-world deployment. In extreme large-scale out-of-core settings, our asynchronous data pipeline successfully scales to \textbf{one billion points} and delivers up to a \textbf{10.5$\times$} end-to-end speedup by effectively overlapping PCIe communication with computation. 
Furthermore, under dynamic shape deployments, our \compileheuristic matches the optimally tuned kernel performance within a negligible \textbf{0.3\%} margin while slashing the compilation and configuration tuning overhead by up to \textbf{175$\times$}. 
Together, these results demonstrate that \method delivers not only theoretical kernel efficiency, but also stable, highly deployable end-to-end acceleration for modern AI workloads.

\section{Related Work}
\lblsect{related-work}

\mypara{The evolution of K-Means workloads.} 
Traditionally, \kmeans was deployed in offline data mining and classical computer vision \cite{1056489, mcqueen1967smc}. However, modern AI pipelines have transformed it into a high-frequency, online primitive across diverse domains. In large-scale data processing and retrieval, \kmeans is heavily utilized for web-scale semantic deduplication \cite{abbas2023semdedup} and embedding quantization for late-interaction search \cite{khattab2020colbert,santhanam2022colbertv2, santhanam2022plaid}. In Large Language Models (LLMs), online clustering is critical for context scaling, enabling dynamic token routing for sparse attention \cite{roy2020routingtransformer, wang2021clusterformer, zhu2025tactic} and semantic state merging for KV cache compression \cite{liu2025clusterkvm, hooper2025kvquant}. More recently, this paradigm has extended to state-of-the-art generative video models, where batched \kmeans is invoked repeatedly during forward passes to perform semantic-aware token permutation in Diffusion Transformers \cite{xi2025sparsevideogen,yang2025sparsevideogen2} and extreme low-bit KV-cache quantization for auto-regressive generation \cite{xi2026quantvideogen}. Consequently, the primary evaluation metric for \kmeans has gradually shifted from offline analytical throughput to strict online invocation latency.

\mypara{Algorithmic optimizations for K-Means.} 
To accelerate the clustering process, prior research has focused on reducing the theoretical computational complexity. Methods leveraging the triangle inequality safely skip redundant distance calculations \cite{10.5555/3041838.3041857, pmlr-v37-ding15}. Other approaches shrink the effective dataset size using summarization or sampling to estimate centroid updates \cite{bachem2018scalablekmeans, 10.1145/1772690.1772862}. Recent theoretical literature also continues to explore dual-distance metrics to improve mathematical convergence \cite{gada2025novelkmeansclusteringapproach}. While these algorithmic innovations successfully decrease arithmetic operations, standard GPU implementations of \kmeans remain fundamentally IO-bound, making implementation-level optimization a critical prerequisite to fully translating algorithmic FLOP reductions into end-to-end wall-clock speedups.

\mypara{Hardware-aware and IO-optimized ML primitives.} 
Recent systems research achieves massive performance breakthroughs by optimizing operator data paths and memory hierarchies rather than altering the underlying mathematical formulas. The most prominent example is FlashAttention \cite{dao2022flashattention, dao2023flashattention2, shah2024flashattention3}, which fuses attention computation to entirely bypass the explicit materialization of the massive $N \times N$ attention matrix in HBM. This IO-aware philosophy has rapidly expanded to yield heavily optimized primitives for dynamic LLM serving \cite{ye2025flashinfer} and memory management \cite{kwon2023vllm}. Furthermore, in handling highly dynamic and skewed data distributions, systems frequently transform irregular scatter writes into sorting followed by regular segmented operations to eliminate severe atomic contention \cite{dao2023flashdecoding, guo2025sonicmoe}. These works collectively demonstrate that restructuring an operator's memory dataflow to respect hardware synchronization and bandwidth limits is the most effective path to accelerating modern AI primitives.

\section{Preliminary and Motivation}
\lblsect{motivation}

\subsection{Lloyd's Algorithm and Standard GPU Implementation}
\lblsect{motivation-baseline}

Given data points $X\in\mathbb{R}^{N\times d}$ and centroids $C\in\mathbb{R}^{K\times d}$, Euclidean \kmeans minimizes the objective:
\begin{equation}
\min_{a\in\{1,\dots,K\}^N,\; C}\;\sum_{i=1}^{N}\left\lVert x_i-c_{a_i}\right\rVert_2^2
\end{equation}

Lloyd's algorithm solves this by alternating between two stages. The first is the \textbf{assignment stage}, which computes a distance matrix $D\in\mathbb{R}^{N\times K}$ and assigns each point to the nearest centroid:
\begin{equation}
D_{ik}=\left\lVert x_i-c_k\right\rVert_2^2,\qquad a_i=\arg\min_{k} D_{ik}
\end{equation}

The second is the \textbf{centroid update stage}, which aggregates the points assigned to each cluster to compute the new centroids:
\begin{equation}
n_k=\sum_{i=1}^{N}\mathbb{I}[a_i=k],\quad s_k=\sum_{i=1}^{N}\mathbb{I}[a_i=k]\,x_i,\quad c_k\leftarrow \frac{s_k}{n_k}
\end{equation}

In practice, the squared distance in the assignment stage is expanded as $\left\lVert x_i-c_k\right\rVert_2^2=\lVert x_i\rVert_2^2+\lVert c_k\rVert_2^2-2x_i^\top c_k$, which enables the reuse of norms and maps the computation to high-throughput matrix multiplications. 

Many existing GPU libraries follow a direct translation of these equations, as outlined in Algorithm~\ref{alg:standard_kmeans}. It first computes and writes the massive distance matrix $D$ to High Bandwidth Memory (HBM), then reads it back for row-wise reduction. Finally, it performs scatter-style atomic additions at the token granularity to accumulate cluster statistics.

\begin{algorithm}[htbp]
\caption{Standard \kmeans implementation (one iteration)}
\label{alg:standard_kmeans}
\KwIn{Data $X\in\mathbb{R}^{N\times d}$, Centroids $C\in\mathbb{R}^{K\times d}$ resident in HBM}
\KwOut{Assignments $a\in\{1,\dots,K\}^N$, Updated Centroids $C_{new}$}
\BlankLine
\tcp{Stage 1: Assignment}
\textbf{Kernel 1 (Distance):} Load blocks of $X$ and $C$ from HBM\;
Compute $D_{ik} = \lVert x_i\rVert_2^2+\lVert c_k\rVert_2^2-2x_i^\top c_k$\; 
Write full matrix $D\in\mathbb{R}^{N\times K}$ to HBM 
\tcp*{Massive intermediate materialization}
\BlankLine
\textbf{Kernel 2 (Argmin):} Load row $D_{i*}$ from HBM, 
Compute $a_i = \arg\min_k D_{ik}$ for each point $i$\;
Write assignments $a \in \{1,\dots,K\}^N$ to HBM\;
\BlankLine
\tcp{Stage 2: Centroid Update}
\textbf{Kernel 3 (Aggregation):} Initialize sum $s \leftarrow 0$, count $n \leftarrow 0$ in HBM\;
\ForEach{point index $i \in \{1,\dots,N\}$ in parallel}{
    \textbf{Read} feature $x_i$ and assignment $a_i$ from HBM\;
    \texttt{atomic\_add}( $s_{a_i}$, $x_i$ ) \tcp*{Severe atomic contention on hot clusters}
    \texttt{atomic\_add}( $n_{a_i}$, $1$ )\;
}
\BlankLine
\textbf{Kernel 4 (Normalization):} \textbf{Read} $s, n$ from HBM\;
Compute $c_k = s_k / n_k$ for each $k$, and Write $C_{new}$ to HBM\;
\end{algorithm}

\subsection{Kernel-Level Bottlenecks in Standard Implementation}
\lblsect{motivation-bottlenecks}

While the standard implementation is mathematically straightforward, our profiling reveals that its end-to-end performance on modern GPUs is severely limited by two low-level bottlenecks: the memory wall in the assignment stage and atomic serialization in the update stage.

\mypara{Distance materialization bottlenecks the assignment stage.} 
The key inefficiency in the assignment stage is the explicit materialization of the distance matrix $D\in\mathbb{R}^{N\times K}$. The matrix $D$ is a short-lived intermediate that is written to HBM by the distance kernel and immediately read back by the assignment kernel. This dataflow introduces at least $\Theta(NK)$ writes plus $\Theta(NK)$ reads per iteration, which heavily dominates the total memory traffic when $N$ and $K$ are large. 

In contrast, reading the inputs $X$ and $C$ only costs $\Theta(Nd+Kd)$, and writing the outputs costs $\Theta(N+Kd)$. Consequently, even if the distance computation utilizes high-throughput primitives, the end-to-end assignment latency is fundamentally bandwidth-limited by the $2\cdot\Theta(NK)$ HBM round trips required to materialize $D$.

\mypara{Scatter-style updates cause severe atomic write contention.} 
In the centroid update stage, the standard implementation relies on a token-to-cluster aggregation view. Each Cooperative Thread Array (CTA) processes a contiguous block of tokens in their original order, reading the feature $x_i$ and its assigned cluster id $a_i$, and then issues scatter-style atomic adds to update the global sum and count buffers.

While this implicitly achieves the necessary gather operation, the write path is highly inefficient. Local token blocks often contain repeated and interleaved cluster ids. As a result, multiple warps frequently attempt to update the same centroid simultaneously, leading to irregular and highly concentrated write destinations. This causes severe atomic contention, serialization, and cache-line thrashing. Since atomics are issued at token granularity, the number of atomic operations scales as $O(Nd)$. In our profiling on an NVIDIA H200 GPU, this write-path contention restricts the standard update implementation to only 50~GB/s effective bandwidth, drastically falling short of the hardware's theoretical reduction bandwidth limits.

\subsection{System-Level Constraints in Real Deployments}
\lblsect{motivation-system}

Beyond the core kernel inefficiencies, \kmeans faces additional system-level constraints when deployed in modern AI workloads. 
First, these workloads are frequently invoked with extremely large batch sizes, causing the input data to exceed the GPU's VRAM capacity. Processing the data in chunks can reduce peak memory usage, but it introduces frequent Host-to-Device (CPU-to-GPU) communication and synchronization, shifting the bottleneck to the PCIe bandwidth. 
Second, real-world AI pipelines exhibit highly dynamic shapes---the number of points, clusters, and feature dimensions can change rapidly between invocations. Standard implementations that rely heavily on shape specialization require exhaustive auto-tuning to achieve optimal performance. Under dynamic workloads, this configuration search triggers frequent recompilations, significantly increasing the \ttfr and hindering practical usability in online scenarios.

\section{Methodology}
\lblsect{method}



In this section, we introduce \method{}, an efficient \kmeans implementation for modern GPU workloads. We first present \FA in \sects{flash-assign}, which fuses distance computation and reduction to eliminate IO overheads. We then present \siu in \sects{argsort-inverse}, which explicitly constructs an inverse mapping to rewrite high-contention atomic scatter into a regular segment-level aggregation. Finally, we discuss several algorithm-system co-design optimizations in \sects{co-design} to improve deployability in real systems.

\subsection{\FA: Materialization-Free Assignment via Online Argmin}
\lblsect{flash-assign}


To eliminate the severe HBM traffic caused by materializing the distance matrix $D\in\mathbb{R}^{N\times K}$ (\sects{motivation-bottlenecks}), we propose \textit{\FA}. \FA fuses the distance computation and row-wise reduction into a single streaming procedure, ensuring that the full $N\times K$ distance matrix is never explicitly constructed in memory.

\mypara{Online argmin.} 
\FA relies on an online argmin update. For each point $x_i$, we maintain two running states in registers: the current minimum distance $m_i$ and the corresponding centroid index $a_i$. We initialize $m_i=+\infty$ and $a_i=-1$, and then scan the centroids in tiles. For each centroid tile, the kernel computes local distances on chip, identifies the local minimum within that tile, and compares it with the running $(m_i, a_i)$ to keep the smaller one. Repeating this update over all centroid tiles yields the exact global minimum.

\mypara{Tiling and asynchronous prefetch.}
\FA uses two-dimensional tiling over both the points and the centroids. For each tile, the kernel maintains the running $(m,a)$ states on chip and scans centroid tiles sequentially. To hide memory latency, we implement double buffer and asynchronous prefetch, ensuring that loading the next centroid tile from HBM overlaps seamlessly with the distance computation of the current tile.

\begin{algorithm}[h]
\caption{FlashAssign (materialization-free assignment)}
\label{alg:flashassign}
\KwIn{Data $X\in\mathbb{R}^{N\times d}$, centroids $C\in\mathbb{R}^{K\times d}$, point tile size $B_N$, centroid tile size $B_K$}
\KwOut{Assignments $a\in\{1,\dots,K\}^{N}$}

Precompute norms $\|x_i\|_2^2$;

\ForEach{point tile $X_{\mathrm{tile}}$ of size $B_N$ in parallel}{
    Initialize on-chip running states: $m \gets +\infty$, $a \gets -1$\;
    Prefetch the first centroid tile $C_{\mathrm{tile}}^{(0)}$ from HBM into on-chip buffer\;
    
    \For{$t \gets 0$ \KwTo $\lceil K/B_K\rceil - 1$}{
        \If{$t+1 < \lceil K/B_K\rceil$}{
            Prefetch $C_{\mathrm{tile}}^{(t+1)}$ into the alternate buffer\;
        }
        Compute local distances between $X_{\mathrm{tile}}$ and $C_{\mathrm{tile}}^{(t)}$ on chip\;
        Compute tile-local minima and indices $(\tilde m, \tilde a)$ for each point in $X_{\mathrm{tile}}$\;
        Update running states using online argmin\;
        $m \gets \min(m, \tilde m)$, and update $a$ with the corresponding index\;
        Swap buffers\;
    }
    
    Write final assignments $a$ for $X_{\mathrm{tile}}$ to HBM\;
}
\end{algorithm}

\mypara{IO complexity.}
By fusing distance computation and reduction, \FA fundamentally changes the dataflow. Under ideal streaming execution, each assignment pass only needs to read the point matrix $X$ and centroid matrix $C$ once, and write the assignment vector $a$ once. This reduces the dominant IO complexity from $O(NK)$ to $O(Nd + Kd)$, entirely removing the $2\cdot \Theta(NK)$ HBM traffic penalty of standard implementations.

\subsection{\SIU: Low-Contention Centroid Aggregation}
\lblsect{argsort-inverse}

\begin{figure*}[t]
    \centering
    \includegraphics[width=\textwidth]{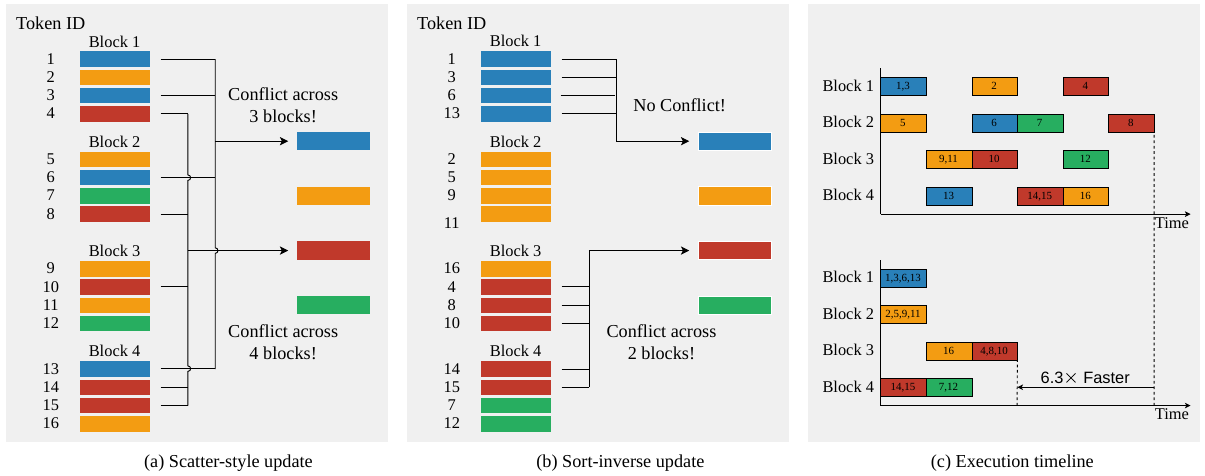} 
    \caption{Illustration of the centroid update stage and atomic contention. 
    \textbf{(a) Standard scatter-style update:} Tokens are directly scattered to their assigned centroids. The highly irregular mapping leads to severe write-side atomic contention when multiple threads update the same centroid simultaneously. 
    \textbf{(b) \Siu:} \method first sorts the tokens by their cluster IDs and constructs an inverse mapping. This transforms the unstructured scatter into regularized, segment-level localized reductions. 
    \textbf{(c) Execution timeline comparison:} The timeline reveals that the standard approach stalls frequently due to atomic lock contention on HBM, whereas \SIU issues contention-free memory writes, significantly hiding latency and accelerating the reduction phase.}
    \label{fig:siu_mechanism}
\end{figure*}

To resolve the severe write-side serialization in the centroid update stage (\sects{motivation-bottlenecks}), we propose \textit{\siu} as illustrated in \fig{siu_mechanism}. Instead of performing scatter-style atomic additions at the token granularity, we explicitly restructure the execution to enable localized, segment-level reductions.

\mypara{Explicit inverse mapping via argsort.}
The core idea is to transform the token-to-cluster update into a cluster-to-token gather. We first apply an \texttt{argsort} operation to the assignment vector $a$ to obtain the permutation index \texttt{sorted\_idx}, and then form the sorted cluster-id sequence $a^{\mathrm{sorted}}$. In this sorted logical order, identical cluster ids naturally group together to form contiguous \emph{cluster-id segments}. Note that this sorting is performed only on the 1D assignment vector $a$; we do \emph{not} physically permute the heavy point matrix $X$ in memory.

\mypara{Segment-level localized aggregation.}
Once the inverse mapping is built, each CTA processes a contiguous chunk of the sorted sequence $a^{\mathrm{sorted}}$. The CTA identifies the cluster-id segments boundaries inside its assigned chunk and uses \texttt{sorted\_idx} to gather the corresponding token features from the original $X$ matrix. 
Crucially, the CTA accumulates partial sums and counts entirely in fast on-chip memory (registers or shared memory) for each segment. It only issues global \texttt{atomic\_add} operations to HBM at segment boundaries. By keeping the read-side gather from HBM but moving the reduction logic on chip, we effectively bypass the write-path bottleneck.
\fig{siu_mechanism}(b) provides an overview of this reorganization.

\begin{algorithm}[h]
\caption{\SIU (low-atomic centroid update)}
\label{alg:sort_inverse_update}
\KwIn{Points $X\in\mathbb{R}^{N\times d}$ (original order), assignments $a\in\{1,\dots,K\}^{N}$, chunk size $B_N$}
\KwOut{Centroid sums $s\in\mathbb{R}^{K\times d}$, counts $n\in\mathbb{R}^{K}$, updated centroids $C$}

\textbf{Sort by cluster id:} compute \texttt{sorted\_idx}$\leftarrow \texttt{argsort}(a)$\;
Construct sorted cluster ids $a^{\mathrm{sorted}}[j] \leftarrow a[\texttt{sorted\_idx}[j]]$\;

Initialize $s \leftarrow 0$, $n \leftarrow 0$\;

\For{$l \gets 0$ \KwTo $N-1$ step $B_N$}{
    $r \leftarrow \min(l + B_N, N)$\;
    Load $a^{\mathrm{sorted}}[l:r]$ and \texttt{sorted\_idx}$[l:r]$\;
    Identify contiguous segments of identical cluster ids in $a^{\mathrm{sorted}}[l:r]$\;

    \For{each segment $(u,v,k)$ in $a^{\mathrm{sorted}}[l:r]$}{
        Gather token features from original order using indices \texttt{sorted\_idx}$[u:v]$\;
        Accumulate local partial sum $\Delta s_k$ and local partial count $\Delta n_k$ on chip\;
        \textbf{Atomic merge (once per segment):} \texttt{atomic\_add}$(s_k, \Delta s_k)$ and \texttt{atomic\_add}$(n_k, \Delta n_k)$\;
    }
}

\For{$k \gets 1$ \KwTo $K$}{
    $c_k \leftarrow s_k / n_k$\;
}
\end{algorithm}


\mypara{Atomic add count analysis.}
\Siu drastically reduces the total number of atomic operations. In a standard scatter update, atomics are issued per token, scaling as $O(Nd)$. In our design, atomics are issued per contiguous cluster-id segment. After sorting, the sequence contains exactly $K$ contiguous segments. Since the workload is partitioned into chunks of size $B_N$ across CTAs, chunk boundaries may split a segment, adding at most $\lceil N/B_N\rceil$ extra boundaries. Therefore, the worst-case number of atomic merges drops to $O((K+\lceil N/B_N\rceil)d)$. This theoretical reduction directly translates to the elimination of write contention.

\subsection{Efficient Algorithm-System Co-design}
\lblsect{co-design}

\mypara{Large-scale data processing via \cso.}
When the input cannot reside entirely in GPU memory and must be staged from CPU memory, we use a \cso design.
We partition the data into chunks and use CUDA streams to coordinate asynchronous host-to-device transfers and \kmeans computation. 
Each chunk is copied to the GPU with non-blocking transfers, then processed by assignment and centroid update kernels, while the next chunk is transferred in parallel, following a double-buffer streaming pattern. 

\mypara{\Compileheuristic for fast \ttfr.}
Configuration tuning in \method{} compilation often introduces a large \ttfr overhead, especially under dynamic shapes and cross-hardware deployment. 
To address this issue, we design a \compileheuristic that selects high-quality kernel configurations directly from hardware characteristics (in particular, L1 and L2 cache sizes) and problem shape, instead of relying on expensive exhaustive tuning. 
The goal is not to perfectly tune every shape, but to reliably match near-best configurations at very low compilation cost. 

\section{Experiments}
\label{sec:experiments}


\begin{figure}[t]
    \centering
    \includegraphics[width=\textwidth]{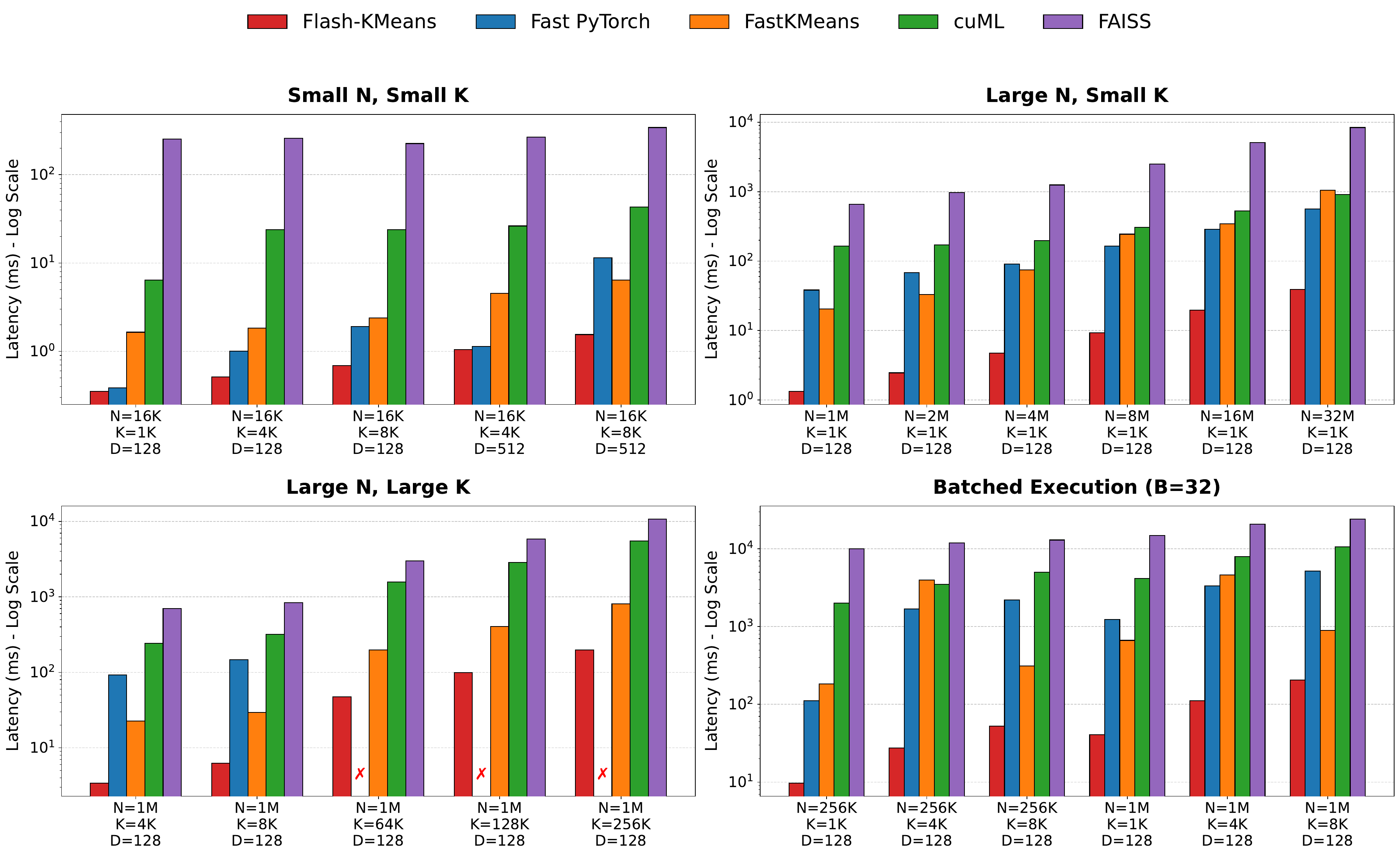}
    \caption{End-to-End Latency of \method compared to standard baselines. We group the evaluation into four representative regimes. The y-axis is presented in log scale to accommodate magnitude differences. Red \textbf{\texttimes} marks denote Out-Of-Memory failures (e.g., standard PyTorch fails to explicitly materialize the $N \times K$ distance matrix in Large $K$ workloads). \method delivers consistent and substantial speedups across diverse shapes and handles extreme limits gracefully.}
    \label{fig:e2e_benchmark}
\end{figure}

\subsection{Experimental Setup}
\lblsect{exp-setup}


We evaluate \method on an NVIDIA H200 GPU with CUDA 12.8. We benchmark Euclidean \kmeans over a comprehensive sweep of data sizes $N$, cluster counts $K$, feature dimensions $D$, and batch sizes $B$. 
We compare \method against four heavily optimized baselines: \texttt{fast\_pytorch\_kmeans}, \texttt{fastkmeans} \cite{fastkmeans2025}, NVIDIA \texttt{cuML} \cite{raschka2020machine}, and \texttt{FAISS} \cite{johnson2019billion}. 
We report end-to-end latency, individual kernel latencies (assignment and centroid update), and \ttfr for dynamic shape deployments.

\subsection{Efficiency Evaluation}
\lblsect{exp-efficiency}

\mypara{End-to-end speedup benchmark.}
\fig{e2e_benchmark} compares the end-to-end latency per iteration across implementations under different workloads. For clarity, we group the results into three representative regimes: large $N$ with large $K$, large $N$ with small $K$, and small $N$ with small $K$. 
In the memory-intensive large-$N$, large-$K$ regime (e.g., $N=1\mathrm{M}, K=64\mathrm{K}, D=512$), standard PyTorch implementations run out of memory due to the massive distance matrix, whereas \method achieves the largest absolute speedup, outperforming the best baseline (\texttt{fastkmeans}) by over \textbf{5.4$\times$}. 
In the compute-intensive large-$N$, small-$K$ regime (e.g., $N=8\mathrm{M}, K=1024$), \method remains the fastest, reducing end-to-end latency by \textbf{94.4\%} (a \textbf{17.9$\times$} speedup over \texttt{fast\_pytorch\_kmeans}). 
In the small-$N$, small-$K$ regime, where kernel launch and framework overheads dominate, \method still delivers consistent acceleration, scaling up to a \textbf{15.3$\times$} speedup in highly batched scenarios (e.g., $B=32$). 
Overall, \method remains robust across batch sizes and feature dimensions, demonstrating that its IO-aware optimizations generalize across diverse shape configurations.

\begin{figure}[t]
    \centering
    \includegraphics[width=\textwidth]{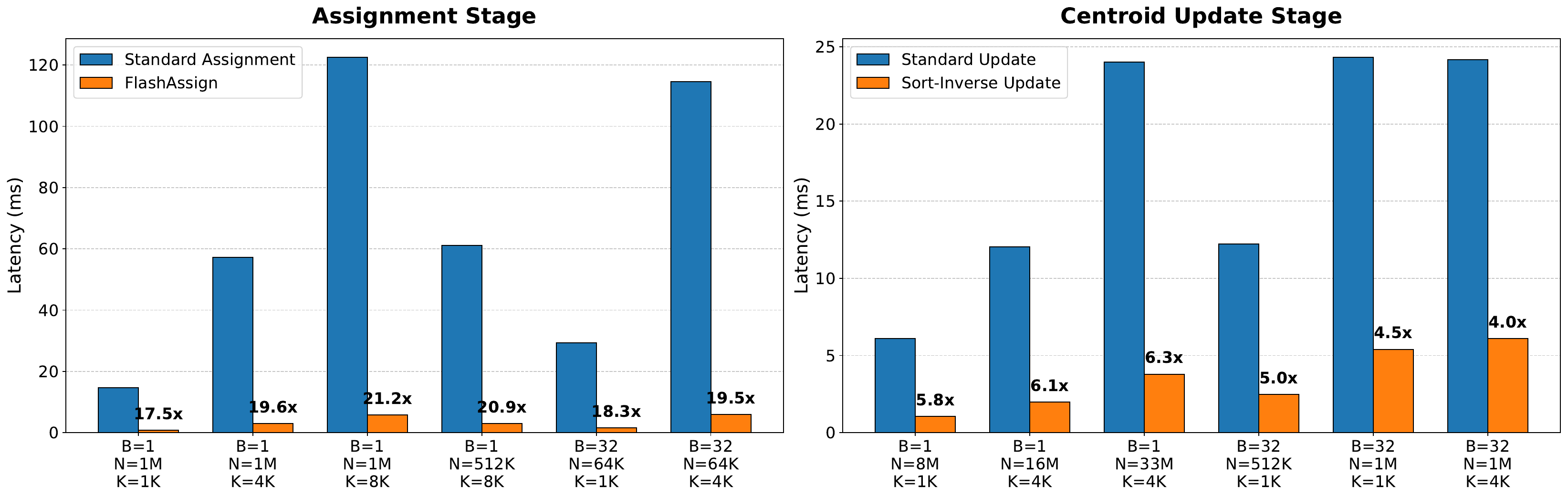}
    \caption{Kernel-level latency breakdown. Latency comparison of our custom kernels versus standard implementations across diverse extreme workloads ($D=128$). \textbf{Left:} \FA completely removes HBM distance materialization, scaling up to a 21.2$\times$ speedup. \textbf{Right:} \SIU replaces per-token scatter atomic adds with sorted localized merges, eliminating atomic contention and achieving up to a 6.3$\times$ speedup.}
    \label{fig:kernel_speedup}
\end{figure}


\mypara{Kernel-level efficiency breakdown.}
To isolate the sources of our end-to-end acceleration, we benchmark the two core \kmeans stages individually against highly optimized standard implementations. \fig{kernel_speedup} presents the absolute latency comparisons and relative speedups across six representative boundary configurations for each stage.
On a highly demanding configuration ($N=1\mathrm{M}, K=8192$), \FA drastically reduces execution time from $122.5\mathrm{ms}$ down to just $5.8\mathrm{ms}$, achieving up to a \textbf{21.2$\times$} speedup over the standard assignment implementation. 
Concurrently, 
\siu systematically accelerates the reduction step, reaching up to a \textbf{6.3$\times$} speedup on massive-scale workloads (e.g., $B=1, N=33\mathrm{M}, K=4096$). 
These microbenchmarks definitively confirm that \method effectively neutralizes both the dominant IO limits and synchronization bottlenecks, aligning directly with our hardware analysis in \sects{motivation-bottlenecks}.

\subsection{Algorithm-System Co-design Evaluation}
\lblsect{exp-codesign}

\begin{figure}[t]
    \centering
    \includegraphics[width=\textwidth]{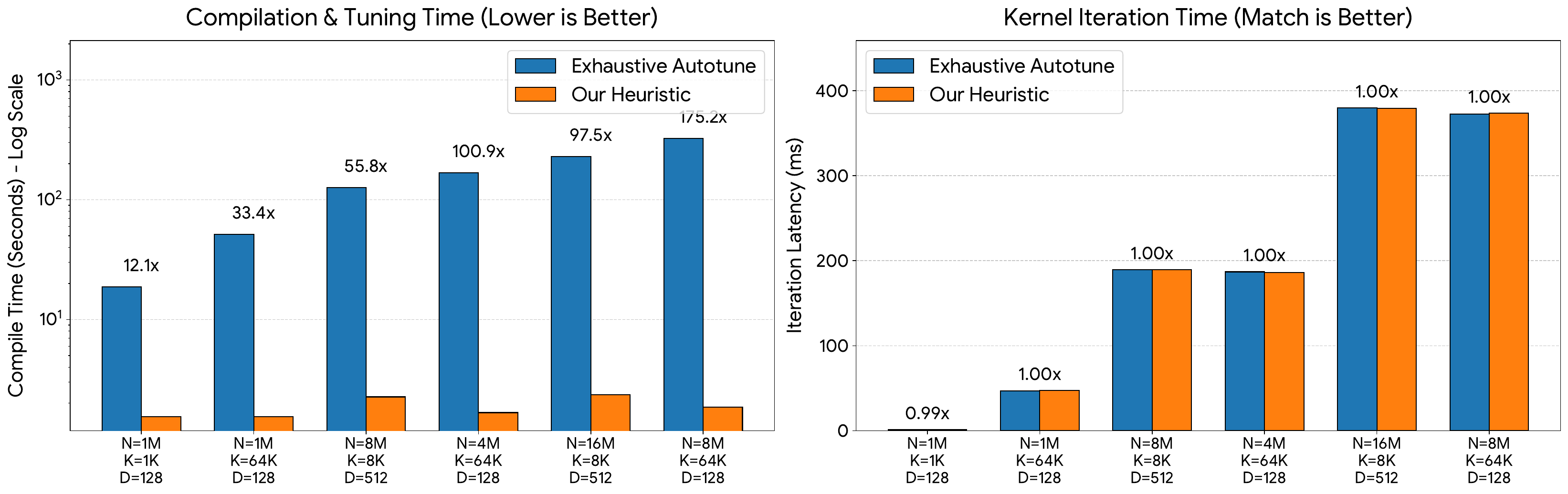}
    \caption{Effectiveness of the \compileheuristic. \textbf{Left:} Compilation and configuration search time (log scale). The heuristic slashes tuning overhead by up to 175$\times$, entirely bypassing the severe compilation bottleneck of exhaustive auto-tuning. \textbf{Right:} Kernel iteration latency. The heuristic seamlessly matches the optimal performance found by exhaustive search across various shapes, ensuring near-zero runtime degradation.}
    \label{fig:compile_benefit}
\end{figure}


\mypara{Large-scale out-of-core data processing.}
To evaluate \method in memory-constrained, massive-scale scenarios, we benchmark workloads where the dataset severely exceeds GPU VRAM capacity, scaling up to one billion points. Under these extreme conditions, standard PyTorch implementations immediately fail due to Out-Of-Memory errors. We therefore compare \method against \texttt{fastkmeans}, the most robust out-of-core baseline available.
On an extreme workload of one billion points ($N=10^9, K=32768, D=128$), \method completes an iteration in just \textbf{41.4 seconds}, whereas the baseline requires 261.8 seconds, yielding a \textbf{6.3$\times$} speedup. Furthermore, on a configuration of $N=400\mathrm{M}$ and $K=16384$, \method achieves an exceptional \textbf{10.5$\times$} end-to-end speedup (8.4s vs. 88.4s). These results demonstrate that our pipelined execution successfully bounds the peak GPU memory footprint while delivering order-of-magnitude acceleration for massive data processing.

\mypara{Fast \ttfr for dynamic shapes.} 
To evaluate our \compileheuristic, \fig{compile_benefit} compares its compilation time and runtime performance against exhaustive tuning across six representative scale-up workloads.
As shown in the left chart, exhaustive tuning time explodes as the problem size scales, requiring over \textbf{325 seconds} to sweep configurations for massive shapes (e.g., $N=8\mathrm{M}, K=65536$). In contrast, our heuristic analytically derives the optimal configuration in less than \textbf{2.5 seconds} across all tested shapes, achieving up to a \textbf{175$\times$} reduction in \ttfr. 
Crucially, as shown in the right chart, this massive reduction in configuration overhead does not compromise execution efficiency. The kernel iteration latency driven by the heuristic consistently matches the exhaustively tuned oracle, with a negligible performance difference of less than \textbf{0.3\%}. This confirms that \method provides highly deployable, out-of-the-box acceleration for dynamic AI pipelines without requiring offline warm-ups.
\section{Conclusion}
\label{sec:conclusion}

In this paper, we introduced \method, a highly optimized, IO-aware \kmeans implementation designed to overcome the severe memory and synchronization bottlenecks of modern GPU workloads. Rather than altering the underlying mathematics, we systematically restructured the execution dataflow through \FA to eliminate massive distance matrix materialization, and \SIU to resolve write-side atomic contention. Coupled with an asynchronous out-of-core data pipeline and shape-aware compile heuristics, \method delivers up to a \textbf{17.9$\times$} end-to-end speedup over bes baselines, and massively outperforms industry standards like \texttt{cuML} and \texttt{FAISS} by \textbf{33$\times$} and over \textbf{200$\times$}, respectively. By gracefully scaling to extreme workloads of one billion points and slashing compilation overheads by \textbf{175$\times$} with near-zero performance degradation, \method provides a robust, mathematically exact, and highly deployable clustering primitive for next-generation generative AI infrastructure.

\clearpage
\newpage
\bibliographystyle{plainnat}
\bibliography{paper}



\end{document}